\documentclass{sigkddExp}

\usepackage{amssymb,amsmath,epsfig,graphics,float,subfigure,balance,xspace,algorithm,algorithmic}
\usepackage{graphicx,url,booktabs,lineno,cite}
\usepackage{epstopdf,color,dsfont,multirow}
\usepackage[table]{xcolor}

\newtheorem{definition}{Definition}

\DeclareMathOperator*{\argmin}{arg\,min}

\colorlet{tablerowcolor}{gray!15}
\newcommand{\rowcol}{\rowcolor{tablerowcolor}} 

\begin{document}

\title{A Survey on Truth Discovery}

\numberofauthors{1} 
\author{\alignauthor Yaliang Li$^1$, Jing Gao$^1$, Chuishi Meng$^1$, Qi Li$^1$, Lu Su$^1$, \\ Bo Zhao$^2$, Wei Fan$^3$, and Jiawei Han$^4$ \\
\affaddr{$^1$SUNY Buffalo, Buffalo, NY USA}
\\ \affaddr{ $^2$LinkedIn, Mountain View, CA USA}
\\ \affaddr{ $^3$Baidu Big Data Lab, Sunnyvale, CA USA}
\\ \affaddr{ $^4$University of Illinois, Urbana, IL USA}
\\\email{$^1$\{yaliangl, jing, chuishim, qli22, lusu\}@buffalo.edu,\\ $^2$bo.zhao.uiuc@gmail.com, $^3$fanwei03@baidu.com, $^4$hanj@illinois.edu}
}
%\date{}
\maketitle
\begin{abstract}
Thanks to information explosion, data for the objects of interest can be collected from increasingly more sources. 
However, for the same object, there usually exist conflicts among the collected multi-source information. 
To tackle this challenge, truth discovery, which integrates multi-source noisy information by estimating the reliability of each source, has emerged as a hot topic.
Several truth discovery methods have been proposed for various scenarios, and they have been successfully applied in diverse application domains.
In this survey, we focus on providing a comprehensive overview of truth discovery methods, and summarizing them from different aspects. We also discuss some future directions of truth discovery research. 
We hope that this survey will promote a better understanding of the current progress on truth discovery, and offer some guidelines on how to apply these approaches in application domains.
\end{abstract}

\section{Introduction}
\label{sec:intro}

% background -- information exploration
In the era of information explosion, data have been woven into every aspect of our lives, and we are continuously generating data through a variety of channels, such as social networks, blogs, discussion forums, crowdsourcing platforms, etc. These data are analyzed at both individual and population levels, by business for aggregating opinions and recommending products, by governments for decision making and security check, and by researchers for discovering new knowledge. In these scenarios, data, even describing the same object or event, can come from a variety of sources.

% background -- noisy information
However, the collected information about the same object may conflict with each other due to errors, missing records, typos, out-of-date data, etc. 
%A recent study \cite{nbc_news_unreliable} reports that $15$ percent of Internet users said they found only a small portion of online information reliable. 
For example, the top search results returned by Google for the query ``the height of Mount Everest'' include ``$29,035$ feet'', ``$29,002$ feet'' and ``$29,029$ feet''. Among these pieces of noisy information, which one is more trustworthy, or represents the true fact? In this and many more similar problems, it is essential to aggregate noisy information about the same set of objects or events collected from various sources to get true facts.

% voting, and source quality
One straightforward approach to eliminate conflicts among multi-source data is to conduct majority voting or averaging. The biggest shortcoming of such voting/averaging approaches is that they assume all the sources are equally reliable. Unfortunately, this assumption may not hold in most cases. Consider the aforementioned ``Mount Everest'' example: Using majority voting, the result ``$29,035$ feet'', which has the highest number of occurrences, will be regarded as the truth. However, in the search results, the information ``$29,029$ feet'' from Wikipedia is the truth. This example reveals that information quality varies a lot among different sources, and the accuracy of aggregated results can be improved by capturing the reliabilities of sources. The challenge is that source reliability is usually unknown \emph{a priori} in practice and has to be inferred from the data.

% truth discovery
In the light of this challenge, the topic of \textbf{truth discovery} \cite{yin_kdd07,Luna_dependence_vldb09,3estimate_wsdm10,jeff_investment_coling10,luna_survey_vldb12,bo_ltm_vldb12,bo_gtm_qdb12,Jeff_latent_www13,luna_explanation_www13,crh_sigmod14,catd_vldb15,li2015evolving} has gained increasing popularity recently due to its ability to estimate source reliability degrees and infer true information. As truth discovery methods usually work without any supervision, the source reliability can only be inferred based on the data. Thus in existing work, the source reliability estimation and truth finding steps are tightly combined through the following principle: The sources that provide true information more often will be assigned higher reliability degrees, and the information that is supported by reliable sources will be regarded as truths. 

With this general principle, truth discovery approaches have been proposed to fit various scenarios, and they make different assumptions about input data, source relations, identified truths, etc. Due to this diversity, it may not be easy for people to compare these approaches and choose an appropriate one for their tasks. In this survey paper, we give a comprehensive overview of truth discovery approaches, summarize them from different aspects, and discuss the key challenges in truth discovery. To the best of our knowledge, this is the first comprehensive survey on truth discovery. We hope that this survey paper can provide a useful resource about truth discovery, help people advance its frontiers, and give some guidelines to apply truth discovery in real-world tasks. 

% impact and applications
Truth discovery plays a prominent part in information age. On one hand we need accurate information more than ever, but on the other hand inconsistent information is inevitable due to the ``\emph{variety}'' feature of big data. The development of truth discovery can benefit many applications in different fields where critical decisions have to be made based on the reliable information extracted from diverse sources. Examples include healthcare \cite{mukherjee2014people}, crowd/social sensing \cite{aggarwal2013social,LWA+11,WKL+12,GDA_rtss_14,wang2015scalable,miao2015cloudpp}, crowdsourcing \cite{yNIPS09,bo_minority_www14,game_iaai14}, information extraction \cite{jiheng_minority_coling14,li2014entity}, knowledge graph construction \cite{luna_knowledge_kdd14,luna_knowledge_vldb14} and so on. These and other applications demonstrate the broader impact of truth discovery on multi-source information integration.

% organization
The rest of this survey is organized as follows: In the next section, we first formally define the truth discovery task. Then the general principle of truth discovery is illustrated through a concrete example, and three popular ways to capture this principle are presented. After that, in Section \ref{sec:aspect}, components of truth discovery are examined from five aspects, which cover most of the truth discovery scenarios considered in the literature. Under these aspects, representative truth discovery approaches are compared in Section \ref{sec:method}. We further discuss some future directions of truth discovery research in Section \ref{sec:challenge} and introduce several applications in Section \ref{sec:app}. In Section \ref{sec:related}, we briefly mention some related areas of truth discovery. Finally, this survey is concluded in Section \ref{sec:conclusion}.
\section{Overview}
\label{sec:overview}

Truth discovery is motivated by the strong need to resolve conflicts among multi-source noisy information, since conflicts are commonly observed in database \cite{BlNa08,BlFe06,DoNa09}, the Web \cite{wu2011framework,MaWu11}, crowdsourced data \cite{yCVPR08,yEMNLP08,yNIPS12}, etc. In contrast to the voting/averaging approaches that treat all information sources equally, truth discovery aims to infer source reliability degrees, by which trustworthy information can be discovered. In this section, after formally defining the task, we discuss the general principle of source reliability estimation and illustrate the principle with an example. Further, three popular ways to capture this principle are presented and compared.

\begin{table*}[tbh]
\centering
%\begin{tiny}
\begin{tabular}{ccccccc}
\toprule
& George & Abraham  & Mahatma  & John & Barack & Franklin  \\
& Washington & Lincoln & Gandhi & Kennedy &  Obama & Roosevelt \\
\midrule
Source $1$ & Virginia & Illinois & Delhi & Texas & Kenya & Georgia \\
Source $2$ & Virginia & Kentucky & Porbandar & Massachusetts & Hawaii & New York  \\
Source $3$ & Maryland & Kentucky & Mumbai & Massachusetts & Kenya & New York \\
\midrule
Majority Voting & Virginia& Kentucky & Delhi & Massachusetts & Kenya & New York \\
Truth Discovery & Virginia & Kentucky & Porbandar & Massachusetts & Hawaii & New York \\
\bottomrule
\end{tabular}
%\end{tiny}
\caption{Illustrative Example}
\label{example_data}
\end{table*}

\subsection{Task Definition}
 
To make the following description clear and consistent, in this section, we introduce some definitions and notations that are used in this survey.
 
\begin{itemize}
\item An \emph{object} $o$ is a thing of interest, a \emph{source} $s$ describes the place where the information about objects can be collected from, and a \emph{value} $v_o^s$ represents the information provided by source $s$ about object $o$.
\item An \emph{observation}, also known as a \emph{record}, is a $3$-tuple that consists of an object, a source, and its provided value. 
%In some work, the term \emph{fact} is adopted to refer to the pair of an object and a candidate value.
\item The identified \emph{truth} for an object $v_o^*$ is the information selected as the most trustworthy one from all possible candidate values about this object. 
\item \emph{Source weight} $w_s$ reflects the probability of source $s$ providing trustworthy information. A higher $w_s$ indicates that source $s$ is more reliable and the information from this source is more likely to be accurate. Note that in this survey, the terms ``source weight'' and ``source reliability degree'' are used interchangeably. 
\end{itemize}
 
Based on these definitions and notations, let's formally define the truth discovery task as following.
\begin{definition}
 For a set of objects $\mathcal{O}$ that we are interested in, related information can be collected from a set of sources $\mathcal{S}$. Our goal is to find the truth $v_o^*$ for each object $o \in \mathcal{O}$ by resolving the conflicts among the information from different sources $\{v_o^s\}_{s \in \mathcal{S}}$. Meanwhile, truth discovery methods estimate source weights $\{w_s\}_{s \in \mathcal{S}}$ that will be used to infer truths.
\end{definition}
 
%Note that in this problem definition, we do not specify the relations among objects $\mathcal{O}$ or sources $\mathcal{S}$. The prior knowledge about truths or source weights is not specified as well. These conditions are application-dependent, which will be discussed in more detail in Section \ref{sec:aspect}.   

\subsection{General Principle of Truth Discovery}

In this section, we discuss the general principle adopted by truth discovery approaches, and then describe three popular ways to model it in practice. After comparing these three ways, a general truth discovery procedure is given.

As mentioned in Section \ref{sec:intro}, the most important feature of truth discovery is its ability to estimate source reliabilities. To identify the trustworthy information (truths), weighted aggregation of the multi-source data is performed based on the estimated source reliabilities. As both source reliabilities and truths are unknown, the general principle of truth discovery works as follows: If a source provides trustworthy information frequently, it will be assigned a high reliability; meanwhile, if one piece of information is supported by sources with high reliabilities, it will have big chance to be selected as truth. 

% example
To better illustrate this principle, consider the example in Table \ref{example_data}. There are three sources providing the birthplace information of six politicians (objects), and the goal is to infer the true birthplace for each politician based on the conflicting multi-source information. 
%We can initialize the source weights using a uniform distribution, and the aggregated results will be identical to the ones given by majority voting. Then based on the current aggregated results, the reliability degree of each source will be updated and Sources $2$ and $3$ get higher weights as they provide more accurate information. In the next round, the information provided by Sources $2$ and $3$ will be counted more as it is supported by more reliable sources. We repeat this procedure to update both source weights and truths, and it converges at the third round. 
The last two rows in Table \ref{example_data} show the identified birthplaces by majority voting and truth discovery respectively.

By comparing the results given by majority voting and truth discovery in Table \ref{example_data}, we can see that truth discovery methods outperform majority voting in the following cases: (1) If a tie case is observed in the information provided by sources for an object, majority voting can only randomly select one value to break the tie because each candidate value receives equal votes. However, truth discovery methods are able to distinguish sources by estimating their reliability degrees. Therefore, they can easily break the tie and output the value obtained by weighted aggregation. In this running example, \emph{Mahatma Gandhi} represents a tie case. (2) More importantly, truth discovery approaches are able to output a minority value as the aggregated result. Let's take a close look at \emph{Barack Obama} in Table \ref{example_data}. The final aggregated result given by truth discovery is ``Hawaii'', which is a minority value provided by only Source $2$ among these three sources. If most sources are unreliable and they provide the same incorrect information (consider the ``Mount Everest'' example in Section \ref{sec:intro}), majority voting has no chance to select the correct value as it is claimed by minority. In contrast, truth discovery can distinguish reliable and unreliable sources by inferring their reliability degrees, so it is able to derive the correct information by conducting weighted aggregation. As a result, truth discovery approaches are labeled as the methods that can discover ``the wisdom of minority'' \cite{bo_minority_www14,jiheng_minority_coling14}.

Next, we discuss three popular ways to incorporate this general principle in truth discovery methods.

\subsubsection{Iterative methods}
In the general principle of truth discovery, the truth computation and source reliability estimation depend on each other. Thus some truth discovery methods \cite{yin_kdd07,Luna_dependence_vldb09,jeff_investment_coling10,3estimate_wsdm10} are designed as iterative procedures, in which the truth computation step and source weight estimation step are iteratively conducted until convergence. 

In the truth computation step, sources' weights are assumed to be fixed. Then the truths $\{v_o^*\}_{o \in \mathcal{O}}$ can be inferred through weighted aggregation such as weighted voting. For example, in Investment \cite{jeff_investment_coling10}, the sources uniformly ``invest'' their reliabilities among their claimed values, and the truths are identified by weighted voting. To be more specific, each candidate value $v$ receives the votes from sources in the following way: 
\begin{equation}
vote(v) = \left(\sum_{s \in \mathcal{S}_v} \frac{w_s}{|\mathcal{V}_s|}\right)^{1.2},
\label{eq:investment1}
\end{equation}
where $\mathcal{S}_v$ is the set of sources that provide this candidate value, and $|\mathcal{V}_s|$ is the number of claims made by source $s$. As the truths are identified by ranking the received votes, the final results $\{v_o^*\}_{o \in \mathcal{O}}$ rely more on the sources with high weights. This follows the principle that the information from reliable sources will be counted more in the aggregation.

In the source weight estimation step, source weights are estimated based on the current identified truths. Let's still take Investment for example. In truth computation step, the sources invest their reliabilities among claimed values, and now, in source weight computation step, they collect credits back from the identified truths as follows:
\begin{equation}
w_s = \sum_{v \in \mathcal{V}_s}\left(vote(v)\cdot \frac{w_s / |\mathcal{V}_s|}{\sum_{s' \in \mathcal{S}_v} w_{s'} / |\mathcal{V}_{s'}|}\right).
\label{eq:investment2}
\end{equation}
That is, each source gets the proportional credits back from all its claimed values. As the received votes of values grow according to a non-linear function (Eq.\ (\ref{eq:investment1})), the trustworthy information will have higher votes and contributes more to the source weight estimation. Thus the sources that provide trustworthy information more often will get more credits back and have higher weights.

\subsubsection{Optimization based methods}
In \cite{crh_sigmod14,game_iaai14,catd_vldb15,li2015evolving}, the general principle of truth discovery is captured through the following optimization formulation:
\begin{equation}\label{eq:opt}
\argmin_{\{w_s\},\{v_o^*\}}\sum_{o \in \mathcal{O}} \sum_{s \in \mathcal{S}} w_s \cdot d(v_o^s, v_o^*),
\end{equation}
where $d(\cdot)$ is a distance function that measures the difference between the information provided by source $s$ and the identified truth. For example, the distance function can be $0$-$1$ loss function for categorical data, and $L^2$-norm can be adopted for continuous data. The objective function measures the weighted distance between the provided information $\{v_o^s\}$ and the identified truths $\{v_o^*\}$. By minimizing this function, the aggregated results $\{v_o^*\}$ will be closer to the information from the sources with high weights. Meanwhile, if a source provides information that is far from the aggregated results, in order to minimize the total loss, it will be assigned a low weight. These ideas exactly follow the general principle of truth discovery. 

In this optimization formulation (Eq.\ (\ref{eq:opt})), two sets of variables, source weights $\{w_s\}$ and identified truths $\{v_o^*\}$, are involved. To derive a solution, coordinate descent \cite{Bertsekas99} can be adopted, in which one set of variables are fixed in order to solve for the other set of variables. This leads to solutions in which the truth computation step and source weight estimation step are iteratively conducted until convergence. This is similar to the iterative methods.

\subsubsection{Probabilistic graphical model based methods}
\begin{figure}[tbh]
\centering
\includegraphics[width=0.36\textwidth]{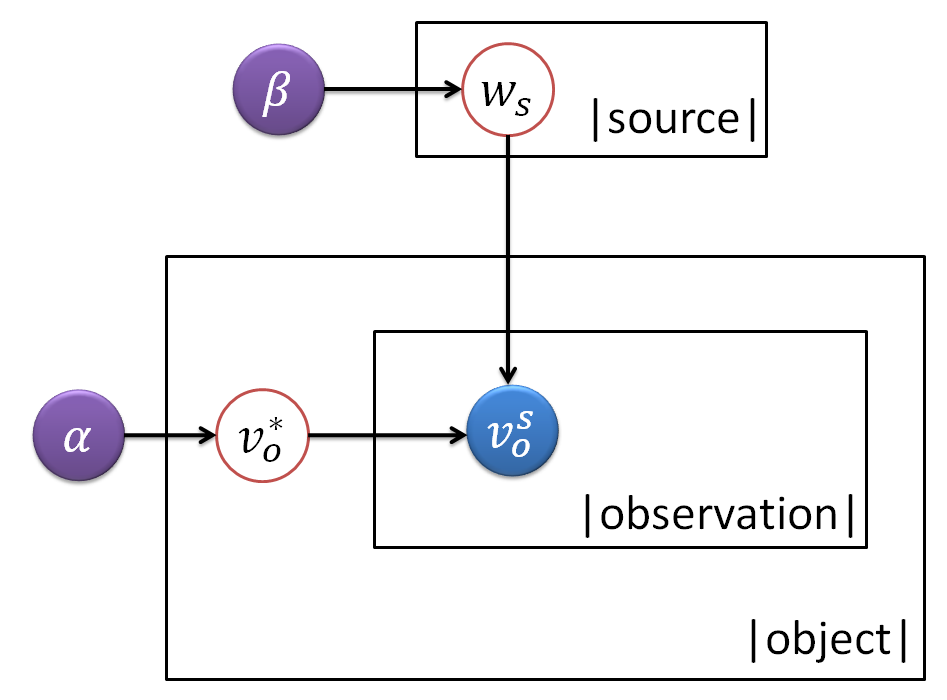}
\caption{The general probabilistic graphical model for truth discovery}
\label{fig_pgm}
\end{figure}

Some truth discovery methods \cite{bo_ltm_vldb12,bo_gtm_qdb12,Jeff_latent_www13} are based on probabilistic graphical models (PGMs). Figure \ref{fig_pgm} shows a general PGM to incorporate the principle of truth discovery, and the corresponding likelihood is following:
\begin{equation}\label{eq:pgm}
\prod_{s \in \mathcal{S}} p(w_s|\beta) \prod_{o \in \mathcal{O}} \left(p(v_o^*|\alpha) \prod_{s \in \mathcal{S}} p(v_o^s|v_o^*,w_s)\right).
\end{equation}

In this model, each claimed value $v_o^s$ is generated based on the corresponding truth $v_o^*$ and source weight $w_s$, and function $p(v_o^s|v_o^*,w_s)$ links them together. Let's take Gaussian distribution as an example. In this case, the truth $v_o^*$ can be set as the distribution mean, while source weight $w_s$ is the precision (the reciprocal of variance). Then the claimed value $v_o^s$ is sampled from this particular distribution with parameter $(v_o^*,w_s)$. If source weight $w_s$ is high, the variance is relative small and thus the ``generated'' claimed value $v_o^s$ will be close to the truth $v_o^*$. In other words, the truth $v_o^*$ is close to the claimed values that are supported by the sources with high weights. Meanwhile, if the claimed value $v_o^s$ is close to the identified truth $v_o^*$, in order to maximize the likelihood, a small variance parameter (high source weight $w_s$) will be estimated. This also reflects the general principle of truth discovery.

To infer the latent variables $\{w_s\}$ and $\{v_o^*\}$, techniques such as Expectation Maximization (EM) can be adopted for inference. The hyperparameters $\alpha$ and $\beta$ also exert their influence on the inference. These parameters can be used to incorporate prior knowledge about truth distribution or source weight distribution.

\subsubsection{Comparison and summary}
Here we compare the above three ways to capture the general principle of truth discovery. 
\begin{itemize}
\item In terms of interpretability, iterative methods are easier to understand and interpret. For example, in Investment and PooledInvestment \cite{jeff_investment_coling10,jeff_generalized_ijcai11}, sources ``invest'' their reliabilities among their claimed values and collect credits back from the identified truths. While optimization based and PGM based solutions are derived from coordinate descent and inference, they are interpretable but need more explanations.
\item When some prior knowledge about sources or objects is available, it can be used to improve the truth discovery performance. In optimization based solutions, prior knowledge can be formulated as extra equality or inequality constraints. For PGM based methods, hyperparameters can capture the external knowledge. 
\end{itemize}
All these three ways are widely used to encode the general principle into truth discovery methods. By providing the above comparison, we do not claim that one of them is better than another. In fact, the coordinate descent in optimization based approaches and the parameter inference in PGM based approaches lead to iterative update rules that are similar to iterative methods, and it is possible to formulate the iterative methods as an optimization problem or a parameter inference task. 

The general procedure of truth discovery is summarized in Algorithm \ref{alg:truth_discovery}. Usually, truth discovery methods start with an initialization of source weights, and then iteratively conduct truth computation step and source weight estimation step. Some stopping criteria are adopted in practice to control the number of iterations. One commonly adopted criterion is to check the change of identified truths or source weights, and terminate the algorithm when the change is smaller than a pre-defined threshold. 

\begin{algorithm}[h]
\caption{\bf General Procedure of Truth Discovery}\label{alg:truth_discovery}
\flushleft
%\begin{scriptsize}
{\bf Input:} Information from sources $\{v_o^s\}_{o \in \mathcal{O}, s \in \mathcal{S}}$. \\
{\bf Output:} Identified truths $\{v_o^*\}_{o \in \mathcal{O}}$ and the estimated source weights $\{w_s\}_{s \in \mathcal{S}}$.
\begin{algorithmic}[1]
\STATE Initialize source weights $\{w_s\}_{s \in \mathcal{S}}$;
\REPEAT
\FOR{$o$ $\leftarrow$ $1$ to $|\mathcal{O}|$}
    \STATE Truth computation: infer the truth for object $o$ based on the current estimation of source weights;
\ENDFOR
\STATE Source weight estimation: update source weights based on the current identified truths;
\UNTIL{Stop criterion is satisfied;}
 \RETURN Identified truths $\{v_o^*\}_{o \in \mathcal{O}}$ and the estimated source weights $\{w_s\}_{s \in \mathcal{S}}$.
\end{algorithmic}
%\end{scriptsize}
\end{algorithm}
\section{Aspects of Truth Discovery}
\label{sec:aspect}
As mentioned above, various truth discovery approaches have been proposed for several scenarios. They have different assumptions about input data, relations among sources and objects, identified truths, etc. Meanwhile, applications in various domains also have their unique characteristics that should be taken into account. These motivate the needs for diverse truth discovery techniques. In this section, we summarize them from the following aspects: input data, source reliability, object, claimed value, and the output.

%%%%%%%%%%%%%%%%%%%%%%%%%%%%%%
%%%%%%%%%%%%%%%%%%%%%%%%%%%%%%
\subsection{Input Data}
We first discuss several features of input data that requires pre-processing conducted before the truth discovery step.

\subsubsection{Duplicate input data} 
It is possible that one source may make several observations about the same object. For example, a Wikipedia contributor may edit the information about the same entry several times, or a crowdsourcing worker can submit his output for a specific task multiple attempts. However, most of the truth discovery methods assume that each source makes at most one observation about an object. If the timestamp for each observation is available, a possible approach is to consider the data freshness \cite{luna_fresh_sigmod14} and select the up-to-date observation. Otherwise, some pre-defined rules can be adopted to choose one from multiple observations.

\subsubsection{Objects without conflict} 
For some objects, all the observations made by sources have the same claimed value. In this case, most of the truth discovery methods should give the same results which is the claimed value (one exception is the method that considers ``unknown'' as an output candidate \cite{jeff_investment_coling10}). So it might be safe to remove these trivial records. Furthermore, in \cite{bo_gtm_qdb12}, the authors report that this pre-processing improves the effectiveness of truth discovery methods. This is because of the fact that if all the sources agree with each other, these observations may not contribute (too much) to the source reliability estimation. It should be pointed out that these trivial records do affect the estimation of source reliability, and thus this pre-processing step should be carefully examined before performing it. 

\subsubsection{Input data format} 
As the information are collected from various sources, they may have different formats \cite{luna_survey_vldb12}. For example, when the object is ``the height of Mount Everest'', some sources have claimed values as ``$29,029$ feet", while others have claimed values as ``$8848$ meters''. Another case, for example, ``John Smith'' and ``Smith, John'', is commonly observed in text data. In fact, these claimed values are the same one and they should be formatted to an identical value.  

\subsubsection{Input uncertainty} 
When the observations are extracted from textual data (for example, in the knowledge fusion task \cite{luna_knowledge_kdd14,luna_knowledge_vldb14}) or the sources provide observations with their confidence indicators (for example, in the question-answering system \cite{game_iaai14}), it is necessary to consider the uncertainty of these observations. In \cite{jeff_generalized_ijcai11}, the authors propose a way to generalize truth discovery methods, which considers multi-dimensional uncertainty, such as the uncertainty in information extractors.

\subsubsection{Structured v.s. unstructured data} 
Previously, most work \cite{yin_kdd07,Luna_dependence_vldb09,3estimate_wsdm10,jeff_investment_coling10} considers the inputs from structured databases. Recently, increasingly more work focuses on unstructured input, such as texts \cite{jiheng_minority_coling14,luna_knowledge_kdd14,ma2015fait}. These unstructured data provide more information such as corpus evidence \cite{jiheng_minority_coling14}, URL, confidence \cite{luna_knowledge_kdd14}, and question text \cite{ma2015fait} which are useful for source reliability estimation. However, at the same time, this extra information introduces more noise and uncertainty. 

\subsubsection{Streaming data}
In many real-world applications, data continue to arrive over time. Most of the existing truth discovery methods are batch algorithms and work on static data. These methods are inefficient to process streaming data as they need to re-run the batch algorithms when new data are available. To tackle this challenge, some truth discovery algorithms \cite{wang2013recursive,zhao2014truth,li2015evolving} have been designed for different types of streaming data.  

\subsubsection{Labeled truths}
Besides the input data, truth discovery methods might assume some additional labeled information. As labeled truths are usually difficult to collect, most truth discovery methods are unsupervised, i.e., estimating the truths without any labeled information. While in \cite{yin_semi_www11,luna_online_vldb11,luna_less_vldb12}, the authors assume that a small set of truths are available and thus the proposed algorithms work in semi-supervised settings. Therefore, a few available truths can be used to guide source reliability estimation and truth computation. The results show that even a small set of labeled truths could improve the performance. 
\vspace{0.1in}

%%%%%%%%%%%%%%%%%%%%%%%%%%%%%%
%%%%%%%%%%%%%%%%%%%%%%%%%%%%%%
\subsection{Source Reliability}
As source reliability estimation is the most important feature of truth discovery, here we examine some assumptions and discuss several advanced topics about the source reliability estimation.

\subsubsection{Source consistency assumption} 
Most truth discovery methods \cite{yin_kdd07,Luna_dependence_vldb09,3estimate_wsdm10,jeff_investment_coling10,yin_semi_www11,bo_gtm_qdb12,crh_sigmod14} have the source consistency assumption, which can be described as follows: A source is likely to provide true information with the same probability for all the objects. This assumption is reasonable in many applications. It is one of the most important assumptions for the estimation of source reliability. 

\subsubsection{Source independence assumption} 
Some truth discovery methods \cite{yin_kdd07,3estimate_wsdm10,jeff_investment_coling10,bo_gtm_qdb12,crh_sigmod14} have the source independence assumption, which can be interpreted as the fact that sources make their observations independently instead of copying from each other. This assumption is equivalent to the following: The true information is more likely to be identical or similar among different sources, and the false information provided by different sources is more likely to be different. 

\subsubsection{Source dependency analysis}
In \cite{Luna_dependence_vldb09,luna_copy_vldb09,luna_dependent_edbt11,luna_copy_vldb10,qi2013group,luna_correlation_sigmod14}, the authors relax source independence assumption. In these work, the authors try to detect copying relationship among sources, and then adjust the sources' weights according to the detected relationship. This copying detection is beneficial in application scenarios where some sources copy information from others. The main principle behind the copy detection is that if some sources make many common mistakes, they are not likely to be independent with each other. However, this principle becomes ineffective when some sources copy information from a good source.

In \cite{Luna_dependence_vldb09}, the authors propose a method to detect direct copying relationship. They assume that a copier does not copy all the information from other sources and may provide some information by itself. In other words, the detected copying relationship is represented by a probability that a source copies the provided information from others. The authors apply Bayesian analysis to infer the existence and direction of copying relationship among sources. This copying detection procedure is tightly combined with truth discovery, and thus the detected copying relationship and the discovered truths are iteratively updated. On the other hand, \cite{qi2013group} alleviates the source dependency problem by revealing the latent group structure among sources, and aggregates the information at the group level. This can reduce the risk of overusing the information from the dependent sources, especially when these sources are unreliable.

Compared with the above work that the copying relationship is detected from snapshots of data, \cite{luna_copy_vldb09} detects copying relationship from dynamic data, where the provided information from sources are changing. The authors apply Hidden Markov Model (HMM) to detect copying relationship given the update history of sources, and the outputs are the evolving copying relationship and the evolving true values. To evaluate the quality of sources, instead of simply considering their accuracy, more metrics are adopted for the dynamic data, such as coverage (how many values in the history a source covers), exactness (how many updates conform to the reality), and freshness (how quickly a source captures a new value). The copying relationship detection is performed by considering the source quality and the update behaviors among sources simultaneously. Similar to \cite{Luna_dependence_vldb09}, both the copying relationship and true values are inferred in an iterative procedure.

However, there may be other complex copying relationships rather than direct copying, such as co-copying (i.e., multiple sources copy from one source), and transitive copying (i.e., a source may copy from other sources transitively). To detect such global copying relationship, \cite{luna_copy_vldb10} extends the idea of \cite{Luna_dependence_vldb09} and \cite{luna_copy_vldb09}: By considering both the completeness and accuracy of sources, the existence and direction of direct copying relationship are detected. Then based on these detected direct copying relationship, Bayesian analysis is performed to detect global copying relationship.

The main principle behind the copy detection is that copiers make the same mistakes with sources they copy from, which is limited in characterizing the relations among sources. In \cite{luna_correlation_sigmod14}, the authors consider different correlations between sources which are more general than copying. Sources may have positive correlations, such as copying information from each other or sharing similar rules for information extractors; sources may also have negative correlations, such as providing data from complementary domains or focusing on different fields of information. Intuitively, the observations from positively correlated sources should not increase the belief that they are true, and the observations supported by only a few negatively correlated sources should not decrease the belief that they are true. The proposed method models correlations among sources with joint precision and joint recall, and Bayesian analysis is preformed to infer the probability of an observation to be true. Approximation methods are also provided as the computation complexity of estimating joint precision and recall is exponential.

In above work, source correlations are inferred from data. While in practice, such source correlation information might be available as extra input. For example, the authors in \cite{wang2014using} fuse the observations from multiple Twitter users by estimating the reliability of each individual user (source). In this application scenario, Twitter users may report observations made by others as their own, which is a form of source copying phenomenon. Instead of estimating the correlations among users, the follower-followee graph and retweeting behaviors in Twitter capture the correlations among users, and such extra information can be adopted to improve the results of truth discovery algorithms.

\subsubsection{Fine-grained source reliability} 
In some scenarios, it is reasonable to estimate multiple source reliabilities for a single source, and thus the variety in source reliability can be captured. The authors in \cite{gupta2011trust} demonstrate that when objects can be clustered into sets, it is better to estimate a source reliability degree for each object set. Similarly, as a website may have very different reliabilities for different attributes of objects, the authors in \cite{yin_semi_www11} treat a website as multiple sources according to attributes and objects, instead of treating the website as a single source. This is equivalent to assigning multiple source reliability degrees to a single website. In \cite{ma2015fait}, the authors propose a probabilistic graphical model to jointly learn the latent topics of questions (objects), the fine-grained source reliability, and the answers to questions (truths). Furthermore, as the number of provided observations becomes very large, the source consistency assumption may not hold any more, and it is more reasonable to have multiple source reliabilities per source. 
\vspace{0.1in}
\subsubsection{Enriched meaning of source reliability} 
In most truth discovery work, source reliability is a parameter that is positively correlated with the probability of a source asserting truths. However in work \cite{bo_ltm_vldb12,catd_vldb15,Jeff_latent_www13,jeff_metrics_asc10,luna_fresh_sigmod14,WKL+12,zhi2015notruth}, the meaning of this parameter is further enriched to fit more complex application scenarios.

In many applications of crowd/social sensing, people want to infer an observation is true or false, for example, whether or not a specific gas station runs out of gas. Thus in \cite{WKL+12}, the authors model the observations from participants as binary variables, and capture the source (participant) reliability by estimating both true positive rate and false positive rate.

In LTM \cite{bo_ltm_vldb12}, the authors assume that the true information about an object can contains more than one value, for example, the authors of a book can be more than one. To estimate the quality of a source, it is insufficient to only calculate the probability that this source's provided information is accurate (precision). Furthermore, the source reliability should also capture the probability that this source fails to provide truths (recall). Thus, in the proposed method LTM, both precision and recall are adopted to estimate the source reliability. On the other hand, it is possible that there is no truth for some objects, e.g., the death date of someone who is still alive. Under this scenario, the authors in \cite{zhi2015notruth} propose several new metrics, such as \emph{silent rate}, \emph{false spoken rate} and \emph{true spoken rate}, to describe the source quality. 

When some sources only provide one or two pieces of information, truth discovery methods may not be able to accurately estimate reliability degrees for these sources. However, such sources are commonly observed and they may contribute to the main part of the collected data. To better deal with such ``small'' sources, in \cite{catd_vldb15}, the authors propose a confidence-aware truth discovery approach that outputs the confidence interval of the source reliability estimation. 

In \cite{jeff_metrics_asc10}, the authors assume that the truths are subjective rather than objective. That is, the identified truths are dependent on the end-users: For different users, the truths can be different. To fit this scenario, instead of representing the reliability of a source as a single scalar, they represent it as three interrelated, but separate values: \emph{truthfulness}, \emph{completeness}, and \emph{bias}. Specifically, truthfulness reflects the probability of a source asserting the truth, completeness reflects the proportion of objects which are covered by a source, and bias is the extent to which a source tends to support a favored position. These three metrics are calculated according to the importance of claims with respect to a specific user, so that they can better reflect users' prior knowledge and preferences.

In \cite{luna_fresh_sigmod14}, the authors study the problem of selecting a subset of sources in dynamic scenarios for truth discovery. To better capture the time-dependent source quality, they define a set of metrics, including accuracy (truthfulness in \cite{jeff_metrics_asc10}), coverage (completeness in \cite{jeff_metrics_asc10}), and freshness (the frequency of updates provided by a source). 

In the probabilistic model of \cite{Jeff_latent_www13}, the source reliability has different semantic meanings in different settings: It can indicate the probability of a source asserting truth, or the probability of a source both knowing and telling the truth, or even the probability of a source intending to tell the truth. The model selection depends on the specific application scenarios.

\subsubsection{Source reliability initialization} 
Most of truth discovery methods start with uniform weights among all sources. As a result, the performance of truth discovery may rely on the majority. When the majority of sources are good, this strategy works well. However, the real scenarios usually are not the case, as sources may copy information from others or they may provide out-of-date information. Nowadays, people apply truth discovery on challenging tasks, such as information extraction and knowledge graph construction. In these challenging tasks, most of the sources are unreliable. For example, \cite{jiheng_minority_coling14} reports that in their task, ``$62$\% of the true responses are produced only by 1 or 2 of the 18 systems (sources)''. This example reveals that a better initialization for source reliability is much in demand. Recent work adopts a subset of labeled data \cite{yin_semi_www11,luna_online_vldb11,luna_less_vldb12}, an external trustful information source \cite{luna_knowledge_kdd14}, or the similarity among sources \cite{jiheng_minority_coling14} as prior knowledge to initialize (or help to initialize) the source reliability.

\subsubsection{Source selection}
We often expect better performance of truth discovery when more data sources are involved in. However, nothing comes for free in practice -- both economic and computational costs should be taken into consideration when applying truth discovery. Moreover, \cite{luna_less_vldb12} shows that the incorporation of bad sources may even hurt the performance of truth discovery. In order to solve these problems, \cite{luna_dependent_edbt11,luna_less_vldb12,luna_fresh_sigmod14} provide methods to wisely select sources for truth discovery constrained by the cost and output quality. 

In \cite{luna_dependent_edbt11}, the authors propose a method to select sources under different cost constraints, in which the task is formulated as an optimization problem. Their Integrating Dependent Sources (IDS) system can return an optimal subset of sources to query, and the selection guarantees that the cost constraints are satisfied. The greedy and randomized approximation algorithms proposed for this problem run in polynomial time with provable quality guarantees.

In \cite{luna_less_vldb12}, source selection is performed on static data sources. The authors first provide a dynamic programming algorithm to infer the accuracy of integrating any arbitrary subset of sources. Then a randomized greedy algorithm is adopted to select the optimal subset of sources by incorporating a source in integration only if the gain of accuracy exceeds the corresponding cost. While in \cite{luna_fresh_sigmod14}, source selection is performed on dynamic data sources whose contents are updated over time. A greedy algorithm is developed to output near-optimal source selection and the optimal frequencies to acquire sources for update-to-date information. 

More interestingly, \cite{bo_minority_www14} claims that a source is useless only when it guesses answers randomly. Even very bad sources can make positive contributions if truth discovery approach assigns negative weights to them. In other words, if a bad source provides a piece of information, we can infer that this information might be wrong with high probability.
%\vspace{0.1in}

%%%%%%%%%%%%%%%%%%%%%%%%%%%%%%
%%%%%%%%%%%%%%%%%%%%%%%%%%%%%%
\subsection{Object}
In this section, we describe how the object difficulty and the relations among objects affect truth discovery.
 
\subsubsection{Object difficulty}
In \cite{3estimate_wsdm10}, 3-Estimates algorithm is proposed, which estimates not only the truth and source reliability, but also the difficulty of getting the truth for each object. This difficulty factor is captured by introducing the trustworthiness of each claimed value. In other words, if a source makes an error on an object, the penalty will be distributed to both the ``source reliability'' factor and the ``object difficulty'' factor. Thus by considering the object difficulty, the errors introduced by objects and sources are separated, and the source reliability can be better estimated. 

\subsubsection{Relations among objects} 
Most of the truth discovery methods assume that objects are independent. In practice, objects may have relations and could affect each other. For example, ``the birth year of a person'' has strong relation with ``the age of a person'', and ``A is the father of B'' indicates that ``B is the son of A''. Such prior knowledge or common sense about object relations could improve the results of truth discovery. 

In \cite{jeff_investment_coling10}, prior knowledge is translated into propositional constraints that are integrated into each round of truth discovery process. Specifically, each fact (an object and its corresponding value) is represented as a $[0,1]$ variable, and then according to the prior knowledge, related facts are combined into propositional constraints. A cost function is defined as the difference between the original results solely based on truth discovery and new results which satisfy the propositional constraints. By minimizing the cost, the probability of each fact to be true is ``corrected'' according to prior knowledge during each iteration. To guarantee the feasibility of this optimization problem, for each object, an augmented ``unknown'' answer is incorporated to relax the constraints and avoid the possible conflicts among constraints.

In \cite{jiheng_minority_coling14}, information extraction and truth discovery methods are combined to solve Slot Filling Validation task which aims to determine the credibility of the output information extracted by different systems from different corpora. The authors construct a Multi-dimensional Truth-Finding Model (MTM) which is a heterogeneous network including systems, corpora, extracted information, and weight matrices between them. Similar to the iterative truth discovery process, the credibility of the extracted information is propagated within the network to infer the reliabilities of both systems and corpora, and in turn, the reliabilities of systems and corpora are used to refine the credibility of the extracted information. To initialize the credibility, the authors consider the dependent relationship among different slots (object) by constructing a knowledge graph. For example, ``Adam is a child of Mary'' should have high credibility if we already believe that ``Bob is a child of Mary'' and ``Adam is a sibling of Bob''.

Temporal and spatial relations also exist among objects. For example, today's high temperature for New York City has correlation with the one of yesterday, and the nearby segments of roads may have correlated traffic conditions. Such temporal and spatial relations among objects can be captured to benefit truth discovery procedure. Recently, \cite{wang2014towards} and \cite{li2015evolving} model the temporal relations among evolving objects on categorical data and continuous data respectively, while \cite{wang2015scalable} and \cite{meng2015truthcorr} can handle both temporal and spatial relations on categorical data and continuous data respectively. Further, \cite{meng2015truthcorr} demonstrates that capturing the relations among objects can greatly improve the performance of truth discovery as objects may receive insufficient observations in many real-world applications.

%%%%%%%%%%%%%%%%%%%%%%%%%%%%%%
%%%%%%%%%%%%%%%%%%%%%%%%%%%%%%
\subsection{Claimed value}
%The data type of claimed value is an important factor to consider when we choose a truth discovery method to apply. Different data types lead to different choices to capture their unique characteristics, such as the concept of distance, the relations among multiple values. 
In this section, we discuss some issues about the claimed values to be considered in the development of truth discovery approaches. 

%\subsubsection{Missing values} 
%Usually, each source only provides observation on a subset of the objects. For the rest of the objects, there are missing values from this source. In some real application scenarios, it is possible and beneficial to fill some missing values by certain assumption. %For example, in one application of social sensing \cite{wang2014using}, people are asked to report the location if they observe that the gas station runs out of gas. Thus, when a person does not report any information, it might indicate that he objects to the claim that ``the gas station runs out of gas''. 

\subsubsection{Complementary vote} 
The complementary vote technique can be used to infer extra information from the claimed values. This technique is based on the single truth assumption that ``there is one and only one true value for each object''. If a source provides a value about an object, \cite{3estimate_wsdm10,bo_ltm_vldb12} assumes that this source votes against other candidate values of this object. This complementary vote is also related with the Local Closed World Assumption (LCWA) \cite{luna_knowledge_kdd14,luna_knowledge_vldb14}, in which any candidate values that violate external knowledge cannot be the truth. The difference between them is that, for LCWA, the trustful external knowledge is used to reject some candidate values of the same object, while for complementary vote, one candidate value is used to reject other candidate values.

\subsubsection{Implication of values} 
In \cite{yin_kdd07}, the concept of implication between claimed values is proposed. For the same object, different claimed values about this object are not independent, instead, they are correlated with each other. Let's consider ``the height of a patient'' as an object, and various sources provide three different claimed values: $175$cm, $176$cm, and $185$cm. If $175$cm is trustworthy, it implies that $176$cm is also trustworthy with a high probability, while the chance of $185$cm being trustworthy is lower. This implication factor between claimed values is also considered in \cite{Luna_dependence_vldb09}.

\subsubsection{Data type of claimed values} 
Actually, the aforementioned implication functions capture the distance of continuous values. In some sense, the complementary vote \cite{3estimate_wsdm10,bo_ltm_vldb12} is based on the distance of categorical values. In fact, it is essential to note that different data types have different concepts of ``distance''. Truth discovery methods, such as \cite{yin_kdd07,Luna_dependence_vldb09,3estimate_wsdm10,jeff_investment_coling10}, focus on categorical data type, though \cite{yin_kdd07,Luna_dependence_vldb09} can be extended to continuous or string data by adding implication functions. On the other hand, GTM \cite{bo_gtm_qdb12} is especially designed for continuous data type. From another perspective, \cite{yin_semi_www11} uses different relationships among observations to capture the property of different data types: for categorical data type, mutual exclusive relations exist among observations, and for continuous data type, mutual supportive relations are modeled. Unfortunately, these relations are slightly limited and usually not easy to establish in practice. Recently, \cite{crh_sigmod14} proposes a general truth discovery framework for heterogeneous data, in which the unique properties of data types are captured by appropriate distance functions. 

\subsubsection{Ambiguity of data type} 
It is interesting to note that data types may be ambiguous sometimes. For example, ``$2$'' can be considered as a continuous number naturally, but under some circumstances, it should be considered as categorical data, such as the class label of an object. If it appears in an address, then it should be considered as a string. This example illustrates that the type of data depends on the specific application, and the assumption about data type should be examined before applying a truth discovery approach.

\subsubsection{Hierarchical structure of claimed values}
As mentioned above, both complementary vote and mutual exclusivity capture the distance of categorical or textual values, but they do not fully explore the distance. In \cite{luna_knowledge_vldb14,luna_knowledge_kdd14}, the authors argue that the claimed values can be represented in a hierarchical value space. For example, Barack Obama's birthplace is \emph{Honolulu}, but it is also true that his birthplace is \emph{Hawaii}, or even the \emph{USA}. In this case, the claimed values are not mutually exclusive, and the complementary vote cannot be applied. It is also different with the multiple truths scenarios, as these true values are linked with each other through a hierarchical structure. 
\vspace{0.1in}

%%%%%%%%%%%%%%%%%%%%%%%%%%%%%%
%%%%%%%%%%%%%%%%%%%%%%%%%%%%%%
\subsection{Output}
In this section, we discuss the following factors that need to be taken into consideration for the outputs of truth discovery: (1) What kind of assumptions we should adopt for the identified truths? (2) Which format of output is better, labeling or scoring? (3) How to evaluate the performance of truth discovery methods? (4) How to interpret the outputs of truth discovery?

\subsubsection{Single truth v.s. multiple truths} 
Most of the truth discovery methods \cite{yin_kdd07,Luna_dependence_vldb09,3estimate_wsdm10,jeff_investment_coling10,bo_gtm_qdb12,crh_sigmod14} hold the ``single truth'' assumption which assumes there is one and only one truth for each object. With this assumption, truth discovery task aims to select the most trustworthy information as truths. In \cite{3estimate_wsdm10, Luna_dependence_vldb09}, this assumption is made to be more explicit and stronger: If a source provides a claimed value for an object, this source is assumed to vote against other possible claimed values for this object.

The ``single truth'' assumption holds in many application scenarios, however it is not always true. For example, there are multiple authors for a book and also multiple actors/actresses for a movie. In \cite{bo_ltm_vldb12}, a probabilistic graphical model LTM is proposed to discover multiple truths for each object. Under such scenarios, only considering source accuracy cannot distinguish the difference between a source with low precision and a source with low recall. However, both precision and recall are essential for discovering multiple truths. Thus LTM considers both false positive and false negative claims, and it can discover multiple truths simultaneously. Similarly, in \cite{luna_correlation_sigmod14}, the authors also calculate both the source precision and recall to discover multiple truths.

\subsubsection{``Unknown'' truths}
In \cite{jeff_investment_coling10}, for each object, an extra candidate output ``unknown'' is considered. This technique is motivated by the following fact: When the relations among objects are represented as constraints, these constraints may contradict each other. Thus it is possible that no solution (truths for all the objects) can satisfy all the constraints. By introducing the output choice of ``unknown'', at least one feasible solution, ``unknown'' for all the objects, is promised. More importantly, for the objects that do not have a truth, the choice of ``unknown'' can be a correct output. For example, it is suitable to give an output of ``unknown'' for the questions (objects) about the death date of someone who is still alive. The authors in \cite{zhi2015notruth} propose a probabilistic model to take into account the truth existence, and the proposed method can assign the output of ``unknown'' to the questions that have no truths.

%More importantly, this ``unknown'' choice can be a possible output when truth discovery methods cannot identify trustworthy information for some objects. Consider the scenario in which the long-tail phenomenon is observed on both source and object sides \cite{luna_knowledge_vldb14}: For some objects, very few sources provide information about them, at the same time, these sources only make claims about these objects. In such cases, as there is no sufficient information to support any claimed values to be the truth, it might be better to allow ``unknown'' outputs.
	
\subsubsection{Labeling v.s. scoring} 
Typically, the outputs of truth discovery methods fall into one of the following two categories: labeling and scoring. The labeling technique assigns a label (true or false) to each claimed value, or assigns a truth to each object. Here note that the truths given by truth discovery methods, such as GTM \cite{bo_gtm_qdb12} and the framework in \cite{jeff_investment_coling10}, are not necessarily be observed or claimed by any sources. The benefit of labeling technique is that the results are ready to use and can be directly fed into other applications, while its drawbacks are that some useful information is lost during the iterations and it does not provide confidence scores for the identified truths. 

This motivates the need for the scoring technique, which assigns a score to each claimed value, usually in the form of probability. Thus the post-processing step is needed to choose the truths by using a cut-off threshold or choosing the top ones for each object. The benefit of the scoring technique is that all the claimed values have certain probability to be chosen as truths and there is less information loss during the iterative procedure. 

%Further in \cite{game_iaai14}, a vectrorize technique is adopted. By applying vectorize technique, the labeling and scoring technique are combined.

\subsubsection{Performance measurement} 
To evaluate the effectiveness of truth discovery methods, various performance metrics are adopted: accuracy (or error rate) for categorical data, Mean of Absolute Error (MAE) and Root of Mean Square Error (RMSE) for continuous data. These metrics are suitable when groundtruths are available. 
%Unfortunately, groundtruths are not always available, especially in tasks such as knowledge graph construction. To evaluate the performance of truth discovery methods, more efforts are much in demand. 
In \cite{yin_semi_www11}, the ``usefulness'' of results is proposed: Not all the results are worth being evaluated, instead, we only care about the useful identified truths. Besides, memory cost and running time can be used to evaluate the efficiency. 

\subsubsection{Output explanation} 
The interpretations of truth discovery results are important for both end-users (``why should I trust the results'') and developers (``how to improve the results''). However, most of the truth discovery methods output opaque information. For example, the output could be that the reliability degree of a source is $6$, or the support from sources for a claimed value is $3.7$. These intermediate parameters are not interpretable for end-users and developers, and we need to construct some intuitive explanations based on the raw outputs.

In \cite{Jeff_latent_www13}, source reliability degree is modeled as the probability of a source asserting the truth, which is more meaningful for human interpretation. In addition, they extend the source reliability definition by considering the difference between knowing the truths and telling the truths. Specifically, in their SimpleLCA model, the reliability is considered as the probability of a source asserting the truths, while in GuessLCA model the reliability is the probability of a source both knowing and asserting the truths, and in MistakeLCA and LieLCA models, the reliability means the probability of a source intending to tell the truths. 

In \cite{luna_explanation_www13}, the authors propose methods to automatically generate both snapshot explanation and comprehensive explanation for truth discovery. These explanations could be used to explain the results of truth discovery derived from iterative computing procedure. Snapshot explanation provides end-users a high-level understanding of the results by gathering all the positive and negative explanations for a specific decision over alternative choices. It lists all the evidences that support the decision as positive explanations and all the evidences which are against the decision as negative ones. In order to make the explanation more concise, categorization and aggregation are conducted so that similar evidences will not appear repeatedly. In addition, they perform evidence list shortening to remove evidence with less importance, and thus human-interpretable explanations can be derived. Compared with snapshot explanation, comprehensive explanation constructs a directed acyclic graph (DAG) to provide an in-depth explanation of the results. The DAG is constructed by tracing the decisions from the last round of iteration all the way to the first iteration where each node represents a decision and the children represent evidences. After the DAG construction, an explanation shortening is performed to eliminate those rounds where no decisions are reached to derive a concise explanation.
\vspace{0.1in}
\section{Truth Discovery Methods}
\label{sec:method}
In the previous section, we summarize various truth discovery methods from five aspects, namely, input data, source reliability, object, claimed value, and the output. Now, we briefly describe several representative truth discovery methods, and compare them under different features (Tables \ref{comparison_methods_1} and \ref{comparison_methods_2}). By providing such a comparison, we hope to give some guidelines so that users and developers can choose an appropriate truth discovery method and apply it to the specific application scenarios.

\begin{table*}[tbh]
\centering
\begin{tabular}{>{\kern-\tabcolsep}*{7}{>$c<$}<{\kern-\tabcolsep}}
\toprule
&\multicolumn{4}{c}{Input Data}  &\multicolumn{2}{c}{Source Reliability} \\
\cmidrule(r){2-5} \cmidrule(r){6-7}
& \multicolumn{1}{c}{Categorical} & \multicolumn{1}{c}{Continuous} & \multicolumn{1}{c}{Heterogeneous} & \multicolumn{1}{c}{Labeled Truth} & \multicolumn{1}{c}{Source Dependency} & \multicolumn{1}{c}{Enriched Meaning}\\
\midrule
\multicolumn{1}{c}{TruthFinder} &  \surd  &  \surd  & & & & \\
\rowcol \multicolumn{1}{c}{AccuSim} & \surd  & \surd  & & & & \\
\multicolumn{1}{c}{AccuCopy} &  \surd  &  \surd  & & &  \surd  & \\
\rowcol \multicolumn{1}{c}{2-Estimates} &  \surd  & & & & & \\
\multicolumn{1}{c}{3-Estimates} &  \surd  & & & & & \\
\rowcol \multicolumn{1}{c}{Investment} &  \surd  & & & & & \\
\multicolumn{1}{c}{SSTF} &  \surd  &  \surd  &  \surd  &  \surd  & & \\
\rowcol \multicolumn{1}{c}{LTM} &  \surd  & & & & &  \surd \\
\multicolumn{1}{c}{GTM} & &  \surd & & & & \\
\rowcol \multicolumn{1}{c}{Regular EM} & \surd &  & & & & \surd \\
\multicolumn{1}{c}{LCA} &  \surd  & & & & &  \surd \\
\rowcol \multicolumn{1}{c}{Apollo-social} & \surd &  & & & \surd & \surd \\
\multicolumn{1}{c}{CRH} &  \surd  & \surd  &  \surd & & & \\
\rowcol \multicolumn{1}{c}{CATD} & &  \surd  & & & &  \surd \\
\bottomrule
\end{tabular}
\caption{Comparison of Truth Discovery Methods: Part 1}
\vspace{0.15in}
\label{comparison_methods_1}
\end{table*}

\begin{table*}[tbh]
\centering
\begin{tabular}{>{\kern-\tabcolsep}*{6}{>$c<$}<{\kern-\tabcolsep}}
\toprule
&\multicolumn{2}{c}{Object} & \multicolumn{1}{c}{Claimed Value} &\multicolumn{2}{c}{Output} \\
\cmidrule(r){2-3} \cmidrule(r){4-4}  \cmidrule(r){5-6}
& \multicolumn{1}{c}{Object Difficulty} & \multicolumn{1}{c}{Object Relation} & \multicolumn{1}{c}{Complementary Vote} & \multicolumn{1}{c}{Multiple Truths} & \multicolumn{1}{c}{``Unknown'' Truths} \\
\midrule
\multicolumn{1}{c}{TruthFinder} & & & & & \\
\rowcol \multicolumn{1}{c}{AccuSim} & & &  \surd  & & \\
\multicolumn{1}{c}{AccuCopy} & & &  \surd  & & \\
\rowcol \multicolumn{1}{c}{2-Estimates} & & &  \surd  & & \\
\multicolumn{1}{c}{3-Estimates} &  \surd  & & \surd  & & \\
\rowcol \multicolumn{1}{c}{Investment} & &  \surd  & & &  \surd  \\
\multicolumn{1}{c}{SSTF} & & &  \surd  & & \\
\rowcol \multicolumn{1}{c}{LTM} & & & &  \surd  & \\
\multicolumn{1}{c}{GTM} & & & & & \\
\rowcol \multicolumn{1}{c}{Regular EM}  & & & & & \\
\multicolumn{1}{c}{LCA} & & & & & \\
\rowcol \multicolumn{1}{c}{Apollo-social}  & & & & & \\
\multicolumn{1}{c}{CRH} & & & & & \\
\rowcol \multicolumn{1}{c}{CATD} & & & & & \\
\bottomrule
\end{tabular}
\caption{Comparison of Truth Discovery Methods: Part 2}
\label{comparison_methods_2}
\end{table*}

Due to space limitation, here we only describe each truth discovery algorithm briefly. For more details about these methods, the readers may refer to the reference papers. 

\begin{itemize}
    \item TruthFinder \cite{yin_kdd07}: In TruthFinder, Bayesian analysis is adopted to iteratively estimate source reliabilities and identify truths. The authors also propose the source consistency assumption and the concept of ``implication'', which are widely adopted in other truth discovery methods.
    \item AccuSim \cite{Luna_dependence_vldb09,luna_survey_vldb12}: AccuSim also applies Bayesian analysis. In order to capture the similarity of claimed values, the implication function is adopted.
	\item AccuCopy \cite{Luna_dependence_vldb09,luna_survey_vldb12}: This method improves AccuSim, and considers the copying relations among sources. The proposed method reduces the weight of a source if it is detected as a copier of other sources.
    \item 2-Estimates \cite{3estimate_wsdm10}: In this approach, the single truth assumption is explored. By assuming that ``there is one and only one true value for each object'', this approach adopts complementary vote.
    \item 3-Estimates \cite{3estimate_wsdm10}: 3-Estimates augments 2-Estimates by considering the difficulty of getting the truth for each object.
    \item Investment \cite{jeff_investment_coling10}: In this approach, a source uniformly ``invests'' its reliability among its claimed values, and the confidence of a claimed value grows according to a non-linear function defined on the sum of invested reliabilities from its providers. Then the sources collect credits back from the confidence of their claimed values. 
    %\item PooledInvestment\cite{jeff_investment_coling10}: PooledInvestment is similar to Investment. The only difference is that the confidence of a claimed value is linearly scaled instead of non-linearly scaled. 
	\item SSTF \cite{yin_semi_www11}: In this semi-supervised truth discovery approach, a small set of labeled truths are incorporated to guide the source reliability estimation. Meanwhile, both mutual exclusivity and mutual support are adopted to capture the relations among claimed values.   
	\item LTM \cite{bo_ltm_vldb12}: LTM is a probabilistic graphical model which considers two types of errors under the scenarios of multiple truths: false positive and false negative. This enables LTM to break source reliability into two parameters, one for false positive error and the other for false negative error.
	\item GTM \cite{bo_gtm_qdb12}: GTM is a Bayesian probabilistic approach especially designed for solving truth discovery problems on continuous data.
    \item Regular EM \cite{WKL+12}: Regular EM is proposed for crowd/social sensing applications, in which the observations provided by humans can be modeled as binary variables. The truth discovery task is formulated as a maximum likelihood estimation problem, and solved by EM algorithm.
    \item LCA \cite{Jeff_latent_www13}: In LCA approach, source reliability is modeled by a set of latent parameters, which can give more informative source reliabilities to end-users. 
    \item Apollo-social \cite{wang2014using}: Apollo-social fuses the information from users on social media platforms such as Twitter. In social network, a claim made by a user can either be originally published by himself or be re-tweeted from other users. Apollo-social models this phenomenon as source dependencies and incorporates such dependency information into the truth discovery procedure. 
    \item CRH \cite{crh_sigmod14}: CRH is a framework that deals with the heterogeneity of data. In this framework, different types of distance functions can be plugged in to capture the characteristics of different data types, and the estimation of source reliability is jointly performed across all the data types together. 
    \item CATD \cite{catd_vldb15}: CATD is motivated by the phenomenon that many sources only provide very few observations. It is not reasonable to give a point estimator for source reliability. Thus in this confidence-aware truth discovery approach, the authors derive the confidence interval for the source reliability estimation.
\end{itemize}

We compare these truth discovery approaches in Tables \ref{comparison_methods_1} and \ref{comparison_methods_2}. Here we summarize them under different features as follows:
\begin{itemize}
    \item \textbf{Input data}: (1) Most of truth discovery methods can handle categorical data. Although GTM and CATD are designed for continuous data, it is feasible to encode the categorical data into probabilistic vectors \cite{game_iaai14}. (2) TruthFinder, AccuSim and AccuCopy can handle continuous data by incorporating the implication function, and SSTF captures the characteristic of continuous data by mutual support. GTM, CRH and CATD are particularly designed for continuous data. (3) Besides CRH that is proposed for heterogeneous data, SSTF can also deal with categorical and continuous data simultaneously. (4) Among these truth discovery methods, only SSTF is proposed to work in semi-supervised setting. However, in practice, most truth discovery methods can be modified to take advantage of labeled truths. One possible solution is to replace some aggregated results in each iteration by corresponding labeled truths, which can thus guide the source reliability estimation.
	\item \textbf{Source reliability}: (1) There is a series of work about source dependency analysis \cite{Luna_dependence_vldb09,luna_copy_vldb09,luna_dependent_edbt11,luna_copy_vldb10,luna_correlation_sigmod14}, and here we only list the AccuCopy algorithm and Apollo-social due to space limitation. (2) LTM, Regular EM, LCA and CATD enrich the meaning of source reliability from different aspects.
	\item \textbf{Object}: Among the discussed truth discovery methods, 3-Estimates is the only one that takes into account the object difficulty. Investment and PooledInvestment approaches can incorporate the relations among objects. More efforts are needed to develop truth discovery methods that consider the advanced features about objects.
    \item \textbf{Claimed value}: Several truth discovery approaches \cite{Luna_dependence_vldb09,luna_survey_vldb12,3estimate_wsdm10,yin_semi_www11} adopt the concept of complementary vote to strengthen the belief in single truth assumption.  	
	\item \textbf{Output}: LTM is proposed to handle the multiple truth scenarios. Investment and PooledInvestment are augmented with ``unknown'' truths. Note that it is possible to modify some truth discovery methods, for example, LTM, to consider the unknown truths by adding one more candidate value ``unknown'' for each object. 
\end{itemize}
\section{Future directions}
\label{sec:challenge}
Although various methods have been proposed, for truth discovery task, there are still many important problems to explore. Here we discuss some future directions of truth discovery.
\begin{itemize}
    \item \textbf{Unstructured data}. For most of the truth discovery approaches, they assume that the inputs are available as structured data. Nowadays, more and more applications of truth discovery are dealing with unstructured data such as text. In \cite{luna_knowledge_vldb14,jiheng_minority_coling14}, the inaccurate information can come from both text corpora and information extractors, i.e., two layers of sources. If the object difficulty is also considered, the penalty of wrong information should be distributed to three factors, corpora, extractors, and objects. This brings new challenges to source weight estimation. Further, the extracted inputs from unstructured data are much more noisy and also bring new information (uncertainty \cite{jeff_investment_coling10}, evidence \cite{jiheng_minority_coling14}, etc.) to be taken into consideration. 
    \item \textbf{Object relations}. For most of the truth discovery approaches, they assume that objects are independent with each other. However, more and more application scenarios reveal the importance of capturing the relations among objects. In knowledge graph \cite{luna_knowledge_kdd14}, objects can be related in various ways, such as ``the birth year of a person'' and ``the age of a person''. Although some efforts \cite{jeff_investment_coling10,wang2015scalable,meng2015truthcorr} have been made to consider the object relations, this problem needs more explorations. A more difficult question is to automatically discover relations among objects. Due to the large scale of the involved objects, it is impossible to manually detect and encode such relations.
    \item \textbf{Source reliability initialization}. Some potential disadvantages are observed on truth discovery with uniform weight initialization: (1) At the beginning of the iterative solutions, the behavior of truth discovery is identical to voting/averaging, and thus truth discovery randomly chooses outputs for tie cases. If the ways of breaking ties are different, the estimated source weights may be quite different. (2) The uniform weight initialization leads to the phenomenon that the aggregated results of truth discovery methods rely on the majority to a certain extent. Nowadays, truth discovery has been applied in more difficult tasks such as slot filling \cite{jiheng_minority_coling14}, in which the majority of sources provide inaccurate information. Thus the performance of truth discovery may suffer from the uniform weight initialization on these tasks. These disadvantages motivate people to investigate new ways to initialize the source reliability. 
    \item \textbf{Model selection}. Given various truth discovery methods, how to select an appropriate one to apply to some specific tasks? Although we compare them from different aspects and give some guidelines in previous section, it is still a difficult task. Is it possible to apply various truth discovery methods together, and then combine the outputs of various methods as final output? This ensemble approach might be a possible solution to tackle the model selection challenge.    
    \item \textbf{Theoretical analysis}. Will the truth discovery methods promise convergence? If so, what is the rate of convergence and is it possible to bound the errors of converged results? These and many more questions need further exploration. Another interesting task is to explore the relations or even the equivalence among various truth discovery approaches. In Section \ref{sec:overview}, we show some evidences that different ways to capture the general principle are equivalent. We may better understand the relations among different truth discovery methods by exploring the equivalence among them.
    \item \textbf{Efficiency}. Most of the existing truth discovery methods adopt iterative procedure to estimate source reliabilities and compute truths. This might be inefficient in practice, especially when we apply truth discovery on large-scale datasets. Some attentions have been given to the efficiency issue in streaming data scenarios \cite{luna_copy_vldb09,wang2013recursive,zhao2014truth,li2015evolving}. However, more efforts are needed to improve the efficiency of truth discovery for general cases.  
    \item \textbf{Performance evaluation}. In the current research work about truth discovery, the groundtruth information is assumed to be available for the purpose of performance evaluation. Unfortunately, in practice, we cannot make this assumption. For example, in the task of knowledge graph construction \cite{luna_knowledge_kdd14}, the number of involved objects is huge and it is impossible to have the groundtruths for performance validation. It requires expensive human efforts to get even a small set of labeled groundtruths. How to evaluate the performance of various truth discovery methods when the groundtruth information is missing? This becomes a big challenge for the truth discovery applications. 
\end{itemize}
\section{Applications}
\label{sec:app}
Nowadays, truth discovery methods have been successfully applied in many real-world applications. 
\begin{itemize}
     
\item \textbf{Healthcare}. In online health communities, people post reviews about various drugs, and this user-generated information is valuable for both patients and physicians. However, the quality of such information is a big issue to address. In \cite{mukherjee2014people}, the authors adopt the idea of truth discovery to automatically find reliable users and identify trustworthy user-generated medical statements. 

\item \textbf{Crowd/social sensing}. Thanks to the explosive growth of online social networks, users can provide observations about physical world for various crowd/social sensing tasks, such as gas shortage report after a disaster, or real-time information summarization of an evolving event. For these participatory crowd/social sensing tasks, users' information may be unreliable. Recently, a series of approaches \cite{LWA+11,WKL+12,aggarwal2013social,wang2014maximum,GDA_rtss_14,wang2015scalable,wang2014towards} have adopted truth discovery to improve the aggregation quality of such noisy sensing data. 

\item \textbf{Crowdsourcing aggregation}. Crowdsourcing platforms such as Amazon Mechanical Turk \cite{url_amazonmturk} provide a cost-efficient way to solicit labels from crowd workers. However, workers' quality are quite diverse, which brings the core task of inferring true labels from the labeling efforts of multiple workers \cite{dawid1979maximum,yNIPS95,yKDD08,yECCV08,yNIPS09,yICML09,yNIPS10,yNIPS12,bo_minority_www14,game_iaai14}. Thus crowdsourcing aggregation approaches focus on learning true labels or answers to certain questions. The main difference between crowdsourcing aggregation and truth discovery is that the former is an active procedure (one can control what and how much data to be generated by workers) while the latter is a passive procedure (one can only choose from available data sources).

\item \textbf{Information extraction}. For information extraction tasks, such as slot filling \cite{jiheng_minority_coling14} and entity profiling \cite{li2014entity}, related data can be collected from various corpora and multiple extractors can be applied to extract desired information. The outputs of different extractors can be conflicting, and thus truth discovery has been incorporated to resolve these conflicts. 

\item \textbf{Knowledge base}. Several knowledge bases, such as Google Knowledge Graph \cite{url_googleknowledgegraph}, Freebase \cite{url_freebase} and YAGO \cite{url_yago}, have been constructed, but they are still far from complete. Truth discovery is a good strategy to automatically identify trustworthy information from the Internet. Existing work \cite{luna_knowledge_kdd14,luna_knowledge_vldb14} has demonstrated the advantages of truth discovery on this challenging task. Meanwhile, the source reliabilities estimated by truth discovery can be used to access the quality of webpages \cite{dong2015knowledge}.  
\end{itemize}

Truth discovery is important to big data and social media analysis where noisy information is inevitable. 
Besides the above applications, truth discovery has great potentials to benefit more applications, such as aggregating GPS data from crowds for traffic control, distilling news from social media, and grading homework automatically for massive open online courses.
%For example, the key idea of trust-enhanced recommendation systems \cite{victor2011trust} is that people tend to accept recommendations from other people they trust, and various recommenders such as moleskiing \cite{avesani2005trust}, Epinions.com \cite{url_epinion} and FilmTrust \cite{golbeck2006filmtrust} have been built based on this idea. However, most trust-enhanced recommenders ask their users to explicitly state users' trust degrees about other users, which becomes inefficient or even impossible when the number of users largely increases. It will be helpful to study how to estimate trust degrees among users by applying the idea of truth discovery in an unsupervised or semi-supervised setting. Other potential applications such as aggregating GPS data from crowds for traffic control, distilling news from social media, and grading homework automatically for massive open online courses can also be benefited by combining with truth discovery.

\section{Related Areas}
\label{sec:related}
 
From a broader view, there are some research topics in the area of \emph{aggregation and fusion} that are relevant to truth discovery. These research topics include multi-view learning/clustering, rank aggregation, sensor data fusion, meta analysis, and ensemble learning. However, the problem settings of these research topics are different from that of truth discovery.
Specifically, 
in multi-view learning/clustering\cite{blum1998combining,bickel2004multi,xu2013survey}, the input is data of various feature sets (views) and the task is to conduct classification or clustering on multiple views jointly.
Rank aggregation \cite{dwork2001rank,lin2010rank} has a different input/output space with truth discovery as it focuses on the aggregation of ranking functions. 
In sensor data fusion \cite{luo2002multisensor,mitchell2007multi}, the source reliability is not considered and usually all the sources are treated indistinguishably. 
In meta analysis \cite{dersimonian1986meta,mark2001practical}, different lab studies are combined via weighted aggregation, but the source weight is derived mostly based on the size of the sample used in each lab study. 
In ensemble learning \cite{Zhou12book,aeni2010ensemble}, the weights of base models are inferred through supervised training whereas truth discovery is unsupervised. 
Compared with these research topics, truth discovery methods aggregate data collected from various sources on the same set of features on the same set of objects, and the goal of truth discovery is to resolve the conflicts among multiple sources by automatically estimating source reliability based on multi-source data only.

Another relevant research area is \emph{information trustworthiness analysis}. 
First, in some trustworthiness scoring systems, various methods have been developed for evaluating the reputation or trustworthiness of users, accounts or nodes in the context of social media platforms \cite{nguyen2009trust,tang2014trust}, or sensor networks \cite{aggarwal2013social,tang2013trustworthiness}. However, these systems focus on the computation of trustworthiness or reliability degrees by considering the content and/or links among sources.
Second, in trust-enhanced recommendation systems \cite{victor2011trust,o2005trust}, the recommendation for one user on one item can be computed by a weighted combination of ratings from other users on this item, where weights incorporate trust between users. In such systems, trust and recommendation are derived separately from different types of data -- Trust is derived based on users' provided trust values about others or the recommendation history information, while recommendation is conducted based on current user-item ratings and the derived trust. 
Compared with these studies, truth discovery tightly combines the process of source reliability estimation and truth computation to discover both of them from multi-source data, and source reliability is usually defined as the probability of a source giving true claims. 

\section{Conclusions}
\label{sec:conclusion}
Motivated by the strong need to resolve conflicts among multi-source data, truth discovery has gained more and more attentions recently. In this survey paper, we have discussed the general principle of truth discovery methods, and provided an overview of current progress on this research topic. Under five general aspects, most of the components of truth discovery have been examined. As the existing truth discovery approaches have different assumptions about input data, constraints, and the output, they have been clearly compared and summarized in Tables \ref{comparison_methods_1} and \ref{comparison_methods_2}. When choosing a truth discovery approach for a particular task, users and developers can refer to this comparison as guidelines.

We have also discussed some future directions of truth discovery research. More efforts are highly in demand to explore the relations among objects, which will greatly benefit the real-world applications such as knowledge graph construction. Furthermore, efficiency issue becomes a bottleneck for the deployment of truth discovery on large-scale data. Besides, how to evaluate the performance or validate the identified truths is a big challenge due to the fact that limited groundtruth is available in practice.

\bibliographystyle{abbrv}
\bibliography{yaliangl} 

\end{document}